\documentstyle[prd,aps,epsf]{revtex}
\begin{document}

\draft
%@twocolumnfalse\endcsname
\preprint{{}}
\title{ Supersymmetric $SU(2)_L\times SU(2)_R\times SU(4)_c$ and 
observable neutron-antineutron oscillation}

\author{Z. Chacko and R. N. Mohapatra }
\address{ Department of Physics, University of Maryland,
College Park, MD-20742, USA.}
\date{February, 1998}
\maketitle
\begin{abstract}
{We show that in a large class of supersymmetric $SU(2)_L\times
SU(2)_R\times SU(4)_c$ models with the see-saw mechanism for neutrino masses
and an R-parity conserving vacuum, there are diquark Higgs bosons with 
masses ($M_{qq}$) near the weak scale even though the 
scale of $SU(2)_R\times SU(4)_c$ symmetry breaking is around $10^{10}$ 
GeV. This happens because these masses ($M_{qq}$) arise out of
higher dimensional operators needed to stabilize the charge conserving 
vacuum in the model. This feature has the interesting implication that
the $\Delta B=2$ processes such as  neutron-anti-neutron
oscillation can have observable rates while at the same time yielding
neutrino masses in the range of current interest.}

\end{abstract} \pacs{\hskip 6 cm  UMD-PP-98-94}

\vskip2pc

\section{ Introduction}

A hallmark of the successful standard model of electroweak interactions is
the automatic conservation of baryon and lepton number, a property obeyed
by all known processese involving elementary particles. Even before the
well-known experimental triumphs of the model, this property was 
recognized as very desirable and appealing. On the other hand, its 
supersymmetric extension, the
minimal supersymmetric standard model (MSSM) which promises to explain two 
of the major unsolved problems of the standard model i.e. the origin and 
stability of the weak scale is plagued by uncontrollable amounts of both 
baryon and lepton number violation, known as R-parity violation. Thus a 
heavy price is paid to understand the symmetry breaking of the standard 
model if one insists on staying within the MSSM.

 A model that preserves both the nice properties of the 
MSSM while at the same time
solving the R-parity violation problem is the supersymmetric left-right
 model (SUSYLR) with the see-saw mechanism for neutrino masses\cite{moh}.
Needless to say that the recent hints for neutrino masses provide an 
extra motivation for studying this model in any case.

A detailed analysis of this model has been the subject of several recent 
papers which explore its
vacuum structure and resulting particle 
spectrum\cite{Kuchi,goran,goran1,chacko}.
Such investigations are essential to establish the viability of the model 
since constraints of supersymmetry are known to seriously alter the 
nature of general field theories compared to their nonsupersymmetric 
versions.
A very important result of these investigations is that the requirement of
electric charge conservation by vacuum imposes stringent constraints on 
the scale of left-right symmetry breaking, $v_R$ (or the $W_R$ scale)
 In a large class of models, essentially two possibilities emerge:
(i) the $W_R$ mass is in the TeV range and R-parity is broken spontaneously
\cite{Kuchi} by the vev of $\tilde{\nu^c}$ or (ii) if R-parity is 
conserved by the vacuum, the $W_R$ scale is above $10^{10}$ 
GeV\cite{goran1,chacko}. In case 
(ii), when the $W_R$ scale is close to its minimum allowed value, there
are light doubly charged bosons and fermions with masses in the 100 GeV 
range. There is a simple group theoretical way to understand this. The
essential point is that the requirement of holomorphy of the superpotential
enhances the global symmetry of the theory (making it bigger than the
gauge symmetry). After
the supersymmetry breaking terms are switched on, the minimum of the
theory violates electric charge forcing one to 
include the nonreormalizable terms in the superpotential. They then lead to
lower limits on the $W_R$ mass following from the lightness of the 
pseudo-Goldstone (PG) states (since $M_{PG}\simeq v^2_R/M$).
Thus the low energy model in these theories is the familiar MSSM with 
automatic 
R-conservation plus massive neutrinos and doubly charged particles.
This provides an experimental way to distinguish the SUSYLR models
from the MSSM.

When the SUSYLR model is embedded into the 
$SU(2)_L\times SU(2)_R\times SU(4)_c$\cite{pati} gauge group with symmetry 
breaking
implemented by the Higgs multiplets suggested in Ref.\cite{marshak},
the arguments leading to the above constraint on the $W_R$ scale carry over
and one has $v_R\equiv M_c\geq 10^{10}$ GeV ($M_c$ being the $SU(4)_c$ 
breaking scale). The enlargement of the gauge 
group however has a new and important physical implication that we study
in this paper. Due to the larger dimensionality of the Higgs multiplets, the 
global symmetry 
of the superpotential becomes larger leading to light doubly colored
fields (or the di-quarks) with masses in the 100 GeV range even though the 
$SU(4)_c$ scale is in the range of $10^{10}$ GeV or so. This result is
sharply different from the corresponding nonsupersymmetric case where
the diquark bosons ``tag'' the $SU(4)_c$ scale and has the following
experimental manifestations.

The existence of diquark Higgs bosons in the $SU(2)_L\times 
SU(2)_R\times SU(4)_c$ was shown in 1980\cite{marshak} to imply 
$\Delta B=2$ processes such as 
neutron-anti-neutron oscillation\cite{marshak,kuzmin} at observable levels 
provided the masses ($M_{qq}$) of diquark fields are in the 10-100 TeV range.
Since the natural scale for $M_{qq}$ in the nonsupersymmetric version of the
model is the $SU(4)_c$ scale, observable $N-\bar N$ oscillation required
$SU(4)_c$ scale $M_c$ to be also in this range. Since $M_c\equiv v_R$ also 
represents
the seesaw scale that determines the neutrino masses, this
case would imply a neutrino mass hierarchy of eV-keV-MeV type which 
though strictly not ruled out is not very favored by current experiments
and by cosmological considerations. 
On the other hand, since in the
supersymmetric $SU(2)_L\times SU(2)_R\times SU(4)_c$ model, some of the 
diquark  masses are light despite a high $SU(4)_c$ scale the neutrino
masses which are connected to the $M_c\equiv v_R$ scale can be in 
the milli-eV to eV range as favored by current data while at the same time
giving $N-\bar N$ oscillation at an observable rate. We believe that this 
result should provide a
new incentive to carry out further experimental search for
$N-\bar N$ oscillation, such as the one proposed by the Oak Ridge 
group\cite{yuri}

\section{ The Model}

The gauge group\cite{pati} of the model is $SU(2)_L\times SU(2)_R\times 
SU(4)_c$ (to be denoted shorthand when needed as $G_{224}$).
The matter fields (i.e. the quarks and leptons) belong to one multiplet
transforming as $\Psi$(2,1, 4) and $\Psi^c$(1,2,$\bar 4$). For the
Higgs sector, we follow the discussion in \cite{marshak} and choose  
 the electroweak Higgs bidoublets transforming as $\phi$(2,2,0) and
the triplets as $\Delta$(3,1,2), $\Delta^c$(1,3,-2), $\overline {\Delta}$
(3,1,-2) and $\overline {\Delta}^c$(1,3,2) of the $SU(2)_L\times 
SU(2)_R\times U(1)_{B-L}\times SU(3)_c$ model embedded into the
$G_{224}$ multiplet $\Delta $(3, 1, 10), $\bar{\Delta}$ (3, 1, 
$\overline{10}$);
$\Delta^c$(1, 3, $\overline{10}$) and $\overline{\Delta^c}$(1, 3, 10).
We will also include a parity odd singlet $S$(1,1,1).   
(The numbers in parenthesis refer to their transformation properties
under $ G_{224} $). 
Let us write down the most general potential involving the above fields 
consistent with the
symmetries. We will then use them to obtain the masses for the doubly colored
fields and show that some of them are pseudo-Goldstone bosons and 
therefore their masses are light. 
In order to account for the possibility that right handed scale is large, we
include, in addition
to the renormalizable interactions, all possible nonrenormalizable
interactions of the $\Delta$'s and $\Delta^c$'s among themselves to
lowest order in 1/$M$  where
M is the scale of new physics above the $v_R$. It could be the Planck
scale or some GUT related scale. In this paper we will vary $M$ between
$10^{15}$ to $10^{18}$ GeV.
 The relevant part of the superpotential is
\begin{eqnarray}
W = if(\Psi^{c^T}\tau_2\Delta^c \Psi^c +\Psi^T \tau_2 \Delta \Psi)+ ( M_0 
+\lambda S) Tr(\Delta^c \overline \Delta^c) + (M_0-\lambda S)Tr(\Delta  
\overline \Delta) +\mu_S S^2 \nonumber \\
 + A {[Tr(\Delta^c \overline {\Delta}^c)]}^2 + B Tr(\Delta^c
\Delta^c) Tr (\overline{\Delta}^c \overline{\Delta}^c)
\end{eqnarray}

In the above equation, the A, B, f, $\lambda$ and $M_0$ are 
parameters of the theory with A and B are of order 1/$M$.
To this one must add the soft supersymmetry breaking terms, which have
mass scale in the range of few hundred GeV's. Since supersymmetry
must remain a good symmetry down to the weak scale, the $F$ terms for all 
the fields must be proportional to $m_{3/2}$, the SUSY breaking parameter.

Before turning to discuss the diquark mass spectrum, we point out a very
crucial property of these models found in Ref.\cite{goran,chacko}
and already alluded to in the introduction. 
The requirement of electric charge conservation by the 
vacuum state implies that one must include the $A$ and $B$ terms given in 
Eq.1. Due to the enhanced global symmetry of the renormalizable part  
of $W$, the model has light charged and/or colored fileds, whose masses 
arise from the $B$ term and are therefore proportional to $v_R^2/M$. 
Since present collider data imply that there are no such particles below
50 to 100 GeV, this enables one to derive a lower limit on 
scale of $v_R$ to be $10^{10}$ GeV for $M= 2\times 10^{18}$ GeV 
and slightly weaker otherwise\cite{goran,chacko}.
In what follows, we will use $10^{10}$ GeV as a generic lower limit on 
$v_R$. 

Using Eq. 1, one can give a group theoretical argument for the existence of
light doubly charged and doubly colored particles in the supersymmetric 
limit as follows. For this purpose let us first
ignore the higher dimensional terms $A$ and $B$ as well as the leptonic
couplings $f$. It is then clear that the superpotential has a
complexified $U(30)$ symmetry (i.e. a $U(30)$ symmetry whose parameters are
taken to be complex) that operates on the $\Delta^c$ and $\bar\Delta^c$
fields. This is due to the holomorphy of the superpotential. After one
component of each of the above fields acquires vev
(and supersymmetry
guarantees that both vev's are parallel), the resulting symmetry is the
complexified $U(29)$. This leaves 118 massless fields. Once we bring in the
D-terms and switch on the gauge fields, 18 of these fields become massive
as a consequence of the Higgs mechanism of supersymmetric theories.
That leaves 100 massless fields in the absence of higher dimensional
terms. In the presence of the higher dimensional operators in the 
superpotential, they lead to 50 complex light fields which consist of
18 $\Delta^c_{qq}$ plus 18 $\bar{\Delta^c_{qq}}$ fields; the two doubly 
charged fields of Ref. \cite{chacko} and 12 leptoquark fields of type
$(u^ce^c+d^c\nu^c)$, $d^ce^c$ and their complex conjugate states.
The detailed analysis of the potential leading to these light fields 
in the presence of soft SUSY breaking is identical
to that given Ref.\cite{chacko}. So we do not repeat it here.
The important point is that their masses arise from
the higher dimensional term $B$ and are given by $v^2_R/M$,
as already mentioned.

In this simplest model with only singlets, the strong coupling becomes 
nonperturbative around $10^{6}$ GeV or so which is much below
the $W_R$ scale of $10^{10}$ GeV or so. We therefore extend the model
in such a way that the strong coupling remains perturbative above the
$v_R$ scale. The simplest way to do this is to add $SU(4)_c$ singlet 
but $SU(2)$ triplet fields (denoted by $\delta$ and $\delta^c$) to the 
model. The parity odd singlet
will lift the left-handed part to the $W_R$ scale and make it 
phenomenologically innocuous at low energies. The resulting theory is 
described by a superpotential given by $W+W'$ with $W$ given above and
\begin{eqnarray}
W'=\lambda'' S(\delta^2-{\delta^c}^2)+M'(\delta^2+{\delta^c}^2)\nonumber \\
+\lambda'(\Delta \delta 
\overline{\Delta}+\Delta^c\delta^c\overline{\Delta}^c) 
\end{eqnarray}
The point of the extra field is that in the absence of the higher 
dimensional terms, this reduces the global symmetry to $U(10)\times SU(2)$
in the righthanded sector. The vevs of $\Delta^c$ and $\delta^c$ break this
group down to $U(9)\times U(1)$. This leaves after gauge symmetry 
breaking  24 real massless states or 12 complex states. They are
easily identified to be the twelve color symmetric diquark states 
$\Delta_{u^cu^c}$ and $\overline{\Delta_{u^cu^c}}$. As before, the 
inclusion
of the same higher dimensional terms in the superpotential gives  
mass of order 100 GeV to the $u^cu^c$ fields for $v_R\simeq 10^{10}$ GeV.
 The remaining
diquark fields have masses of order of $<\delta^c>$. We will choose
the tree level parameters of the potential such that $<\delta^c>
\simeq (10^{-3}-10^{-2}) v_R$ in the following discussion. 

An alternative possibility is to include $SU(4)_c$ singlet but 
$SU(2)$ quintet fields (denoted here as $\Sigma \oplus \Sigma^{c}$). 
\begin{eqnarray}
W''=M''(\Sigma^c\Sigma^c+\Sigma\Sigma) 
+\lambda''(\Delta^c\Sigma^c\bar{\Delta}^c+\Delta\Sigma\bar{\Delta}) 
+\lambda''S(\Sigma^2-{\Sigma^c}^2)
\end{eqnarray}
The light particle counting in this case is more subtle since all terms
in the superpotential do not take part in determining the vacuum state.
By explicit calculation we have checked that the particles that are light
in this case are: $\Delta_{u^cu^c}, \overline{\Delta_{u^cu^c}}$, 
$\Delta_{d^cd^c}, \overline{\Delta_{d^cd^c}}$ and 
$\Delta_{d^ce^c}, \overline{\Delta_{d^ce^c}}$. It is
easily checked that their masses come entirely from the higher dimensional
terms in the superpotential. In this case also, the strong coupling becomes
nonperturbative below $v_R$.

\bigskip
\noindent{\bf Neutron-Anti-neutron oscillation}
\bigskip

To see how $N-\bar{N}$ oscillation arises in the various models described 
above, let us include
in the superpotential the following higher dimensional terms involving
the $\Delta^c$ fields:
\begin{eqnarray}
W'=\frac{\lambda_2}{M}\epsilon^{pqrs}\epsilon^{p'q'r's'}
\Delta^c_{pp'}\Delta^c_{qq'}\Delta^c_{rr'}\Delta^c_{ss'}
+\Delta^c\rightarrow \Delta +~~ terms~~ involving~~ 
\overline{\Delta^c}              
\end{eqnarray}
The $SU(2)$ indices have been suppressed for brevity.
We have scaled the nonrenormalizable terms by the same 
scale, M used earlier. So in making estimates
for the $\Delta B=2$ amplitudes, we will vary this scale between the 
two values of $10^{15}$ to $10^{18}$ GeV. Now note that
in conjunction with the $\Delta^c$ mass and coupling terms in the 
superpotential 
$W$, this gives rise to a four scalar $\Delta^c$ coupling with strength 
$\lambda_{eff}$ to be estimated below. As noted in 
Ref.\cite{marshak}, the diagram in Fig. 1 leads to the six quark
$\Delta B=2$ coupling $u^cu^cd^cd^cd^cd^c$ with a strength
\begin{eqnarray}
G_{\Delta B=2}\simeq \frac{\lambda_{eff} 
v_Rf^3}{M^4_{d^cd^c}M^2_{u^cu^c}}
\end{eqnarray}
There are also diagrams involving the exchange of two $u^cd^c$ type Higgs 
bosons in combination with one $d^cd^c$ boson. These are suppressed compared
to the diagram in Figure 1 since the $M_{u^cd^c}\sim v_R$. 
In order to estimate $G_{\Delta B=2}$, we need to know the value of
$\lambda_{eff}$. This will depend on whether we are considering the triplet
or the quintet case.

\noindent{\it (i) Triplet case:}

This case is the most interesting since all the gauge couplings remain 
perturbative until $v_R$ and we therefore discuss it first.
From the superpotential of the model it is easy to see that, 
\begin{eqnarray}
\lambda_{eff}=\lambda_2 <M_0+ \lambda S-\lambda'\delta^c>/ M
\end{eqnarray}
whereas the F-term condition gives the equation for exact supersymmetry
below $v_R$ to be
\begin{eqnarray}
M_0+\lambda S +\lambda'\delta^c=0
\end{eqnarray}
The change in the sign of the coefficient of the $\delta^c$ term is due to
the fact that $\Delta_{u^cu^c}$ and $\Delta_{d^cd^c}$ have opposite $I_{3R}$.
Thus we find that
\begin{eqnarray}
\lambda_{eff}M\equiv\lambda_2 (M_0+\lambda S-\lambda'<\delta^c>)\simeq 
<\delta^c> 
\end{eqnarray} 
From this we estimate $\lambda_{eff}\simeq 10^{-11}-10^{-7}$ depending on
whether we choose the nonrenormalizable term to be scaled by $M_{Pl}$ or
$M_{U}$. 

Taking $M_{u^cu^c}\simeq 100$ GeV, $M_{d^cd^c}\approx <\delta^c>\simeq 
10^{-3}v_R$, we get $G_{\Delta B=2}\simeq (10^{-30}-10^{-33})f^3$
GeV$^{-5}$. To convert this into a $N-\bar N$ transition amplitude
$\delta m_{N-\bar N}$, one must multiply it with the hadronization 
factor\cite{pasu} usually estimated by various methods to be around 
$10^{-4}$ GeV$^6$. This leads to an neutron-anti-neutron oscillation time
equal to $\tau_{N-\bar N}\approx 6\times (10^9-10^{12})$ sec. where we have
chosen $f\approx 1$. On the other hand, if we chose $<\delta^c>\simeq
10^{-2}v_R$, then we would have $\tau_{N-\bar N}\simeq 6\times (10^{12}
-10^{15})$ sec. These estimates for $\tau_{N-\bar N}$ will go down by a 
factor of $\epsilon^3$ if we assume $M_{qq}\sim \epsilon <\delta^c>$. We 
thus see that for plausible values of parameters of
the theory, one can obtain observable $N-\bar N$ transition times. We 
find it very encouraging that we get numbers within the
observable range of a recently proposed experiment at Oak Ridge\cite{yuri}.

\noindent{\it (ii) Quintet case}

This case has the drawback that the strong coupling
becomes nonperturbative below the $v_R$ scale. If we however ignore this
point, observable $\tau_{N-\bar N}$ comes out more easily in this case,
since both $\Delta_{u^cu^c}$ and $\Delta_{d^cd^c}$ are in the 100-1000 GeV 
range. In this case, $\lambda_{eff} M\simeq <M_0+\lambda S + 
\lambda'\Sigma_{00}>$. It therefore vanishes in the supersymmetric limit
and is of order $m_{3/2}$ after soft SUSY breaking terms are included.
We then get
$\lambda_{eff}\simeq \frac{m_{3/2}}{M}$
Now taking $M_{u^cu^c}\approx M_{d^cd^c}\simeq 1000$ GeV and 
$m_{3/2}\simeq 1000$ GeV, we get
\begin{eqnarray}
G_{\Delta B = 2}\simeq (10^{-20}-10^{-23})f^3~~GeV^{-5}
\end{eqnarray}
Choosing $f\approx .01$, we get $\tau_{N-\bar{N}}\simeq 3\times 10^{5}-
3\times 10^{8}$ sec. (using the hadronic factor to be  $10^{-4}$
GeV$^6$) again in the observable range.

Let us end with a few comments: 

(i) In general, the quark couplings to the diquark fields can lead to
flavor changing neutral currents. The point is that the
$f$ coupling connects to all generations; as a result if we denote $a,b$
as the generation indices, then the $\Delta S=2 $ amplitude are induced by
$\Delta_{d^cd^c}$ exchange at the tree level. However, in the triplet model, 
the diquark fields of $d^cd^c$ type are naturally superheavy. As a result,
 there are no dangerous tree level flavor changing neutral currents.
On the other hand, in the quintet model, the $d^cd^c$ diquark fields are
light. We therefore have to resort to fine tuning such as $f_{12}=0$
and $f_{11}=f_{22}$ to prevent large flavor changing neutral currents. 

(ii) Furthermore, our conclusion is independent of the way supersymmetry
is broken in the hidden sector i.e. whether it is gravity or gauge
mediated. Again the arguments for the gauge mediated case are similar
to the ones given in \cite{chacko}.

In conclusion, we have found that in a class of simple supersymmetric
$SU(2)_L\times SU(2)_R\times SU(4)_c$ models, even though the $v_R$ scale is 
dictated by supersymmetry to be near or above $10^{10}$ GeV, some 
of the sextet diquark fields are forced to be light (in the 100 GeV range). 
The presence of these diquark fields can lead to observable 
neutron-anti-neutron oscillation while at the same time 
allowing neutrino masses to be in the currently favored eV range.

\bigskip
\noindent{\Large \bf{Acknowledgements}}
\bigskip

This work is supported by the National Science Foundation 
grant no. PHY-9421385. We thank B. Brahmachari for help in drawing the 
figure.

\begin{figure}[htb]
\begin{center}
\epsfxsize=8.5cm
\epsfysize=8.5cm
\mbox{\hskip -1.0in}\epsfbox{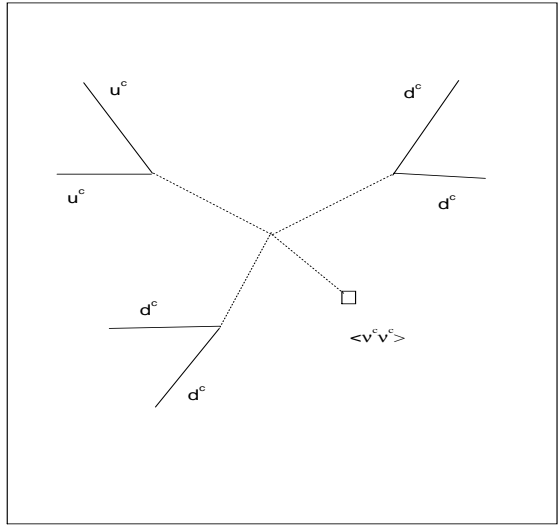}
\caption{ The Feynman diagram responsible for $N-\bar N$ 
oscillation. The unlabelled dashed lines are the scalar diquark 
bosons with appropriate quantum numbers.\label{Fig.1}} \end{center} 
\end{figure}

\end{document}